\documentclass[12pt]{iopart}
\usepackage{iopams}  
\usepackage{graphicx}
\newcommand{\real}{\mathbb{ R}}
\newcommand{\rot}{\mathrm{curl\,}} 
\newcommand{\parcial}[2]{\frac{\partial#1}{\partial#2}} 

\newcommand{\pe}[2]{\left\langle #1,#2\right\rangle}
\begin{document}

\title{Flux through a M\"obius strip?}

\author{L. Fern\'andez-Jambrina}

\address{Matem\'atica e Inform\'atica Aplicadas, E.T.S.I. Navales, Universidad
Polit\'ecnica de Madrid,\\
Avenida de la Memoria 4, \\ E-28040 Madrid, Spain}
\ead{leonardo.fernandez@upm.es}
\vspace{10pt}

\begin{abstract}
Integral theorems such as Stokes' and Gauss' are fundamental in many
parts of Physics.  For instance, Faraday's law allows computing the
induced electric current on a closed circuit in terms of the variation
of the flux of a magnetic field across the surface spanned by the
circuit.  The key point for applying Stokes' theorem is that this
surface must be orientable.  Many students wonder what happens to the
flux through a surface when this is not orientable, as it happens with
a M\"obius strip.  On an orientable surface one can compute the flux
of a solenoidal field using Stokes' theorem in terms of the
circulation of the vector potential of the field along the oriented
boundary of the surface.  But this cannot be done if the surface is
not orientable, though in principle this quantity could be measured on
a laboratory.  For instance, checking the induced electric current on
a circuit along the boundary of a surface if the field is a variable
magnetic field.  We shall see that the answer to this puzzle is simple
and the problem lies in the question rather than in the answer.
\end{abstract}
%
%
%
%
%

\section{Introduction}

The M\"obius strip \cite{pickover} has attracted the interest of
researchers and academics due to its fascinating geometric properties
\cite{macho}.  In spite of its name \cite{mactutor}, it was not discovered
first by August Ferdinand M\"obius, but independently by Johann Benedict Listing \cite{mactutor1},
the father of modern topology.

The construction is fairly simple: starting with a rectangular piece
of paper, one can join two opposite edges in order to form a cylinder.
But if before joining the opposite edges we twist the rectangle 
$180^{\mathrm{o}}$ we obtain this reknowned one-sided surface.

The strip is a non-orientable surface and for this reason it does not
have an outer and an inner side as usual surfaces, such as the sphere,
the plane or the cylinder.  It is a one-sided surface and this fact
has suggested many applications in engineering \cite{macho}.  For
instance, for designing audio and film tapes which could record
longer, since they could be used on the only, but double length, side.
For the same reason, it has been used in printing tapes for printers
and old typewriters.  There are also M\"obius' strips in luggage
conveyor belts in airports in order to double their useful life.  And
a resistor with this shape was patented \cite{resistor, davies}, made
up of two conductive layers and filled with a dielectric material,
preventing residual self-inductance.  There are even aromatic
molecules in organic chemistry with this shape \cite{flapan}.  And we
cannot forget that it is part of the universal recycling symbol,
formed by three green arrows.

But besides academics and engineers, the M\"obius strip has attracted
the attention of many science students.  Just check for ``flux across
a M\"obius strip'' and you will obtain thousands of results in your
favourite search engine.  We focus on their interest in this issue as
target for this paper, as well as their teachers'.

\begin{figure}[h]
		\begin{center}
		    \includegraphics[height=0.2\textheight]{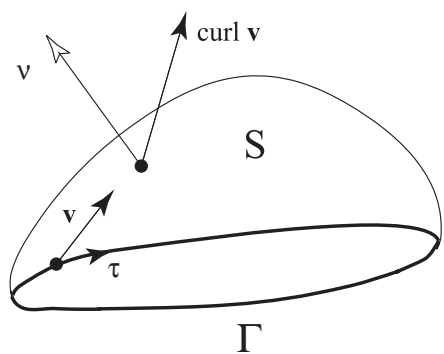}
		\caption{Stokes' theorem\label{stokes}: We have a surface $S$
		with unitary normal $\nu$, bounded by a closed curve $\Gamma$
		with tangent field $\tau$.  We may compute the flux of 
		the vector field $\rot\mathbf{v}$ either by summing up the 
		contributions of $\pe{\rot\mathbf{v}}{\nu}$ on the surface $S$ or 
		the contributions of $\pe{\mathbf{v}}{\tau}$ along the curve 
		$\Gamma$}
		\end{center}
	\end{figure}

The reason for this is that integral theorems such as Stokes' just 
can be applied to orientable surfaces \cite{marsden}, relating the 
flux of the curl of a vector field across a surface with its 
circulation along the boundary of the surface (see Fig~\ref{stokes}).
	
One might think this is a tricky question, since the answer is 
negative: it just cannot be calculated. But there are experiments in 
Physics where one could think this question could have a meaning.

Consider for instance a circuit attached to the boundary of a M\"obius
strip.  According to Faraday's law, the flux of a variable magnetic
field across the surface induces an electric current on the circuit.
One can measure the electromotiv force on the circuit, but in
principle Faraday's law cannot be applied to calculate it with the 
flux across the surface. This is the issue we would like to clarify 
in this paper.

But before providing a solution to this puzzle, we need to recall 
some useful concepts. In Section~2 we review the 
concepts of flux and circulation before stating Stokes' theorem. In 
Section~3 we describe the M\"obius strip as a non-orientable surface. 
As it was expected, the calculations performed on the M\"obius strip 
and on its boundary do not coincide, as Stokes' theorem is not 
applicable, as we show in Section~4. But a simple solution to this issue 
is provided in Section~5. A final section of conclusions is incliuded 
at the end of the paper.

\section{Stokes' theorem}

Before recalling Stokes' theorem, there are a few definitions we 
need to recall: the circulation of a vector field along a curve and 
the flux of a vector field across a surface. This can be reviewed in 
your favourite Vector Calculus book. I have chosen \cite{marsden} for 
its nice examples relating Physics and Mathematics.

The line integral or circulation of a vector field along a curve is 
the generalisation of the concept of the work done by a force along a 
trajectory.

Let us consider a continuous vector field $\mathbf{v}$ and a curve
$\Gamma$ oriented by its tangent field of velocities $\tau$: that is, 
we specify if the curve is followed onwards or backwards.  If the
points on the curve $\Gamma$ are parametrised by 
$\gamma(t)=\left(x(t),y(t),z(t)\right)$,
$t\in[a,b]$, the velocity of this parametrisation is given by
$\tau\big(\gamma(t)\big)=\gamma'(t)$, where the $'$ denotes derivation
with respect to time $t$.  

We define the \textbf{line integral or circulation of $\mathbf{v}$ along $\Gamma$} as
the sum of the projections of $\mathbf{v}$ along $\tau$ at the points on
the curve,
\begin{equation}
\mathcal{C}_{\mathbf{v},\Gamma}:=
\int_{\Gamma}\pe{\mathbf{v}}{\frac{\tau}{\|\tau\|}}\,ds=
\int_a^b\langle \mathbf{v},\tau\rangle_{\gamma(t)}\,
dt,
\end{equation}
taking into account that the length element of a parametrised curve is
$ds=\|\gamma'(t)\|dt$.  The $\pe{}{}$ stands for the scalar or inner
product, whereas $\|\ \|$ stands for the length of a vector.
\begin{figure}[h]
		\begin{center}
		    \includegraphics[height=0.2\textheight]{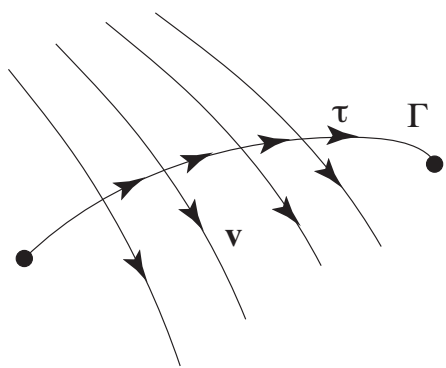}
\caption{Circulation of the vector field $\mathbf{v}$ along the curve 
$\Gamma$\label{circul}: The circulation of the field 
$\mathbf{v}$ along the curve $\Gamma$ is calculated summing up the 
contributions of $\pe{\mathbf{v}}{\tau}$ along the curve $\Gamma$}		\end{center}
\end{figure}

We see that this definition does not change on 
changing the parametrisation of the curve, but it depends on the 
orientation of the curve. That is, it is the same no matter how fast 
we follow the curve. But if we follow the curve the other way round, 
the circulation changes by a sign. (see Fig~\ref{circul}).


On the other hand, the flux integral of a vector field across a
surface is also suggested by examples in Mechanics, Electromagnetism 
and Fluid Mechanics \cite{aris}: the flux of a gravitational field across a 
closed surface is related to the mass contained inside, the flux of a 
electrostatic field is related to the total charge inside the surface 
and the variation of the flux of a magnetic field across a surface is 
related to the electromotiv force induced on the boundary of the 
surface.

Let us consider a compact surface $S$ and a continous vector
field $\mathbf{v}$.  The orientation of the surface is given by a
continuous unitary vector field $\nu$ normal to $S$ at every point. 
For a closed surface, we have just two choices: a vector field pointing 
inwards or outwards.
If such a vector field exists, the surface is called
\textbf{orientable}.  The \textbf{flux of $\mathbf{v}$ across $S$} is
defined as the sum of the projections of $\mathbf{v}$ along $\nu$ at 
the points of the surface,
\begin{equation} \Phi_{\mathbf{v},S}:=\int_S\langle
\mathbf{v},\nu\rangle \,dS,\end{equation}
where $dS$ is the area element of the surface.


If the surface is closed, the orientation of the surface is taken as
positive when $\nu$ points out of the surface.  For a closed surface
then, the flux is positive if more field lines go out of the surface
than enter the surface.

\begin{figure}[h]
\begin{center}
\includegraphics[height=0.2\textheight]{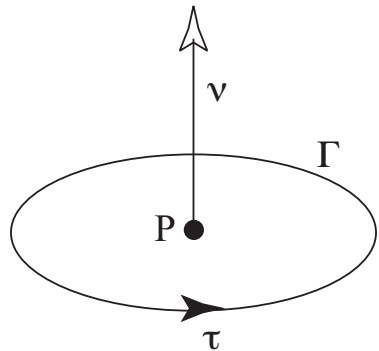}
\caption{Orientation of circunference $S^1$ induced by the one on the 
disk $D^2$\label{orient}: If we set our right thumb along the normal 
$\nu$ to the disk, our fingers show the orientation for $\tau$ along 
the boundary curve $\Gamma$}
\end{center}
\end{figure}

If the surface is open, we can choose either orientation for it.  But
the chosen orientation for $S$ induces an orientation for its boundary
$\Gamma$, as we see in Figs.~\ref{stokes} and \ref{orient}: if our right thumb points as 
the normal vector $\nu$, our fingers show the way the boundary 
$\Gamma$ is to be followed. This convention is necessary to avoid 
amibiguities on stating Stokes' theorem.

For explicit calculations, we usually need a parametrisation for the
points on the surface $S$.  This is a function, with certain 
restrictions \cite{carmo},
$g:D\in\real^{2}\to \real^{3}$, such that
$g(u,v)=\left(x(u,v),y(u,v),z(u,v)\right)\in S$.  That is, we describe
the points of $S$ using curvilinear coordinates $u,v$.

The lines of constant $u$, parametrised by $g(u_{0},v)$ and the lines 
of constant $v$, parametrised by $g(u,v_{0})$, are called coordinate lines of the 
parametrisation $g$ of $S$. Since these lines are contained on the 
surface, their velocities,
\[\mathbf{X_{u}}(u,v)=\frac{\partial g(u,v)}{\partial u},\qquad 
\mathbf{X_{v}}(u,v)=\frac{\partial g(u,v)}{\partial v},\]are tangent 
vector fields to the surface $S$ and their vector product 
$\mathbf{X_{u}}\times \mathbf{X_{v}}$ defines a normal vector field 
to the surface $S$. Hence, a unitary normal vector field is
\[\nu(u,v)=\frac{\mathbf{X_{u}}\times 
\mathbf{X_{v}}}{\|\mathbf{X_{u}}\times \mathbf{X_{v}}\|},\]
but we could have chosen the opposite one, just exchanging the order 
of the coordinates.

If the unitary normal vector field is provided this way, since the 
surface element in such parametrisation is
\[dS=\|\mathbf{\mathbf{X}_{u}}\times\mathbf{\mathbf{X}_{v}}\|\,du\, 
dv,\]
the flux may be computed as
\[
\Phi_{\mathbf{v},S}=\int_{D}\langle \mathbf{v},
\mathbf{\mathbf{X}_{u}}\times
\mathbf{\mathbf{X}_{v}}\rangle\, du\,dv=
\int_{D}\left|
\begin{array}{ccc}
    v^x & v^y & v^z  \\
\frac{\partial x(u,v)}{\partial u} &  \frac{\partial y(u,v)}{\partial u} &
 \frac{\partial z(u,v)}{\partial u}  \\
    \\
\frac{\partial x(u,v)}{\partial v} &  \frac{\partial y(u,v)}{\partial 
v} & \frac{\partial z(u,v)}{\partial v}
\end{array}
\right|_{g(u,v)}du\,dv.\]

It can be seen that this expression is independent of the chosen 
parametrisation, except for the sign due to the choice of orientation.

For instance, a sphere of radius $R$ can be parametrised using the 
colatitude angle $\theta$ and the azimuthal angle $\phi$,
\[g(\theta,\phi)=(R\sin\theta\cos\phi, R\sin\theta\sin\phi, 
R\cos\theta),\quad \theta\in(0,\pi),\ \phi\in(0,2\pi),\]
with some degeneracy, since $g(0,\phi)=(0,0,R)$ is the North pole of the sphere for all 
values of $\phi$ and $g(\pi,\phi)=(0,0,-R)$ is the South pole of the sphere for all 
values of $\phi$.

The lines of constant $\theta$, parametrised by 
$g(\theta_{0},\phi)$, are the parallels of the sphere and 
the lines of constant $\phi$, parametrised by $g(\theta,\phi_{0})$, are the meridians of the sphere.

Now we are ready to state Stokes' theorem. 
Integral theorems such as Green's, Gauss' and Stokes' theorems are
fundamental in Physics, mainly in Fluid Mechanics and
Electromagnetism, since they relate integrals of a field in a region
with integrals on its boundary.  In this sense, they may be viewed as
a way to reduce the dimensions of the integral, but the physical
consequences are far deeper.  This is most relevant for conservative and
solenoidal fields, which can be written respectively as the gradient or the curl of
a potential.

In this paper we are interested in Stokes' theorem, which relates the 
flux integral of the curl of a vector field across a surface with the 
circulation of the field along the boundary of the surface. It may be 
stated as follows:

\textbf{Stokes' theorem:} Let $S$ be a smooth, 
compact, oriented surface, bounded by a curve $\Gamma$. Let 
\textbf{v} be a smooth vector field. The flux of the curl of $\textbf{v}$ across 
$S$, $\Phi_{\rot \mathbf{v},S}$ and the circulation or $\mathbf{v}$ 
along $\Gamma$, $\mathcal{C}_{\mathbf{v},\Gamma}$ are related by\begin{equation}
\Phi_{\rot \mathbf{v},S}=\mathcal{C}_{\mathbf{v},\Gamma}.
\end{equation}
where the orientation for $\Gamma$ is the one induced by the 
orientation of $S$.

The curl is a differential vector operator,
\[\rot \mathbf{v}=\left|
\begin{array}{ccc} \mathbf{e_{x}} & \mathbf{e_{y}} & \mathbf{e_{x}} \\
\partial_{x} & \partial_{y} & \partial_{z}  \\
v^{x} & v^{y} & v^{z} \end{array} \right|,\]
for a vector field
$\mathbf{v}=v^{x}\mathbf{e_{x}}+v^{y}\mathbf{e_{y}}+
v^{z}\mathbf{e_{z}}$ with coordinates $(v^{x},v^{y},v^{z})$ in the
orthonormal trihedron
$\{\mathbf{e_{x}},\mathbf{e_{y}},\mathbf{e_{z}}\}$ of unitary vectors
along the respective axes $X$, $Y$, $Z$.

Stokes' theorem provides a nice interpretation for the curl of a vector field 
$\mathbf{v}$ at a point $P$. Let us consider a small disk $D^{2}$, 
bounded by a circumference $S^{1}$ of radius 
$\varepsilon$ centered at $P$ with unitary normal $\nu$ parallel to 
$\rot \mathbf{v}(P)$ (see Fig.~\ref{orient}).

At lowest order, if the radius $\varepsilon$ is small, we can take
$\rot\mathbf{v}$ as constant on the disk,
\[ \mathcal{C}_{\mathbf{v},S^1(\varepsilon)}=
\Phi_{\rot\mathbf{v},D^2(\varepsilon)}\approx\pi\varepsilon^2\,\|\rot\mathbf{v}(x_{0},y_{0},z_{0}))\|
,\] 
and so we may view the curl of $\mathbf{v}$ at a point $P$ as the
density of circulation of this field on the orthogonal plane, since
\[ \|\rot\mathbf{v}(P))\|=\lim_{\varepsilon\to 0}\frac{\mathcal{C}_{\mathbf{v},S^1(\varepsilon)}}
{\pi\varepsilon^2}.\] 

Hence, the curl of a field shows the existence of closed field lines
or whirlpools (finite circulation) around a point.  Besides, its
direction provides the orientation of these whirlpools. This is 
related to the fact that solenoidal fields are generated by currents 
instead of charges.

One typical example of application of Stokes' theorem is Faraday's 
law, one of Maxwell's laws for Electromagnetism \cite{marsden}, which relates the 
electrical field  $\mathbf{E}$ with the magnetic field
$\mathbf{B}$ through
\begin{equation}\rot \mathbf{E}=-\parcial{\mathbf{B}}{t}.\end{equation}

If we calculate the circulation of the electric field along a closed 
curve $\Gamma$, after applying Stokes' theorem to a surface $S$ 
bounded by $\Gamma$, we get
\[\mathcal{C}_{\mathbf{E},\Gamma}=\Phi_{\rot\mathbf{E},S}=
-\Phi_{\parcial{\mathbf{B}}{t},S}=-\parcial{\Phi_{\mathbf{B},S}}{t},\]
using Faraday's law and taking out the derivative with respect to time.

If we think of the curve $\Gamma$ as a closed circuit, the
circulation of $\mathbf{E}$ is the electromotive force induced by the 
varying magnetic field. This is the simple principle which explains how 
electric motors work.

Another useful application of the theorem is the calculation of the 
flux of a solenoidal field $\mathbf{v}=\rot \mathbf{A}$, that is, of 
a vector field $\mathbf{v}$ endowed with a vector potential 
$\mathbf{A}$,
\begin{equation}
\Phi_{\mathbf{v},S}=\mathcal{C}_{\mathbf{A},\Gamma},\end{equation}
so that it equals the circulation of its vector potential along the 
boundary of the surface. 

According to this result, the flux of the solenoidal field
$\mathbf{v}$ does not depend on the surface $S$, but just on its
boundary $\Gamma$.  If the surface is closed, there is no boundary and
the flux of a solenoidal field across closed surfaces is always zero.
For open surfaces, the flux is the same across \emph{any} other surface bounded by
$\Gamma$.  This fact shall be useful for our purposes later on.

\section{M\"obius strip}
\begin{figure}[h]
		\begin{center}
		    \includegraphics[height=0.2\textheight]{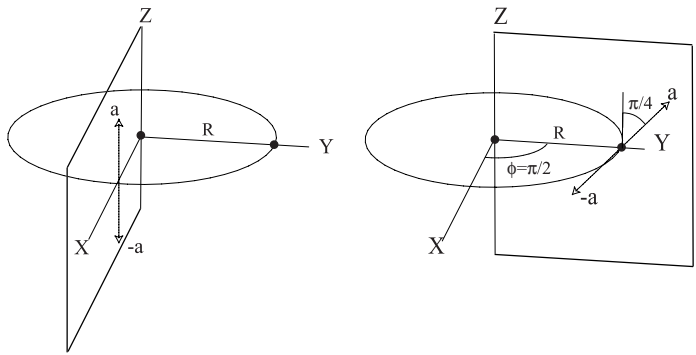}
\caption{Initial location of the segment $I$ and after its center 
rotates $\phi=\pi/2$\label{mob1}}		\end{center}
\end{figure}

\begin{figure}[h]
		\begin{center}
		    \includegraphics[height=0.2\textheight]{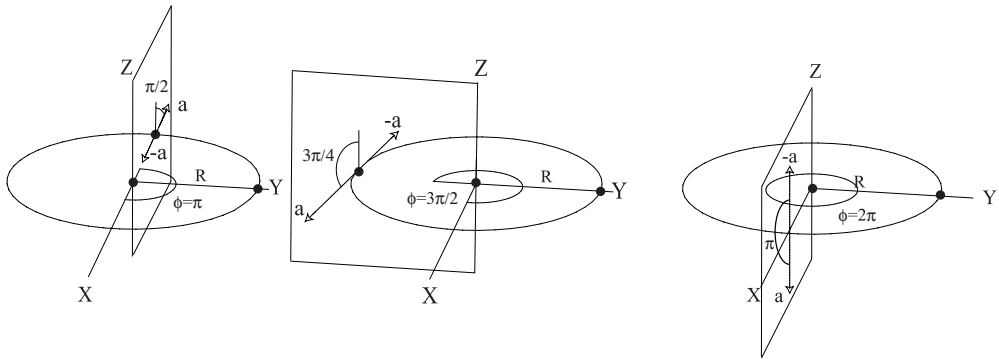}
\caption{Location of the segment $I$ after its center rotates $\phi=\pi, 
3\pi/2,2\pi$\label{mob2}}
\end{center}
\end{figure}

\begin{figure}[h]
		\begin{center}
		    \includegraphics[height=0.2\textheight]{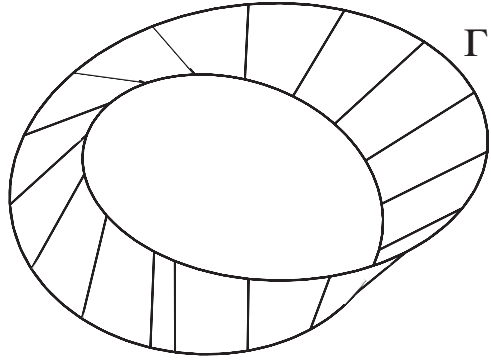}
\caption{M\"obius strip\label{moebius}}
\end{center}
\end{figure}

As we mentioned in Section~1, building a M\"obius strip is
fairly simple (see, for instance, page 106 in \cite{carmo}).  Let us 
consider a vertical segment
$I=\{(R,0,z): z\in [-a,a]\}$ of length $2a$ and the circumference $C$
of radius $R>a$ and center $(0,0,0)$, lying on the plane $z=0$. If we 
rotate the segment $I$, keeping it vertical, along the 
circumference $C$, we would obtain a circular cylinder. But we allow 
the segment also to rotate upside down on travelling along $C$ in 
such a way that the segment is always contained in the plane 
described by the $Z$ axis and the radius of the circumference through 
the center of the segment (see Figs.~\ref{mob1} and \ref{mob2}).
The resulting surface $S$ is a M\"obius strip (see Fig.~\ref{moebius}), which may be 
parameterised in a simple fashion with such a geometric construction.

It seems reasonable to use as parameters the position of a point on
the segment $I$, $r\in(-a,a)$ and the angle $\phi$ rotated by the 
center of the segment along the circumference. 

When the center of the segment has rotated and angle $\phi$ along the
circumference, the segment rotates an angle $\phi/2$ around its center.  If the center
of the segment had not rotated along the circumference, it would have
been parametrised as $(R+r\sin(\phi/2),0,r\cos(\phi/2))$.  But since
it has rotated an angle $\phi$ along the circumference, we have
\[
g(r,\phi)=\left
(\left(R+r\sin\frac{\phi}{2}\right)\cos\phi,
\left(R+r\sin\frac{\phi}{2}\right)\sin\phi,
r\cos\frac{\phi}{2}\right),
\]
for $r\in(-a,a)$, $\phi\in(0,2\pi)$ as a parametrisation for the M\"obius' strip.

That is, $g(r,\phi)$ describes the position of the original point
corresponding to $r\in(-a,a)$ after rotation of the segment by an
angle $\phi/2$ and rotation of its center along the circumference by 
an angle $\phi$.

Using the velocities of the coordinate lines,
\begin{eqnarray*}
\mathbf{X_{r}}(r,\phi)&=&\left(\sin\frac{\phi}{2}\cos\phi, 
\sin\frac{\phi}{2}\sin\phi, \cos\frac{\phi}{2}\right),\\
\mathbf{X_{\phi}}(r,\phi)&=&\left
(-\left(R+r\sin\frac{\phi}{2}\right)\sin\phi,
\left(R+r\sin\frac{\phi}{2}\right)\cos\phi,
0\right)\\&+&\frac{1}{2}\left(
r\cos\frac{\phi}{2}\cos\phi,
r\cos\frac{\phi}{2}\sin\phi,-
r\sin\frac{\phi}{2}\right),\end{eqnarray*}
we may obtain a normal vector field, $\mathbf{X_{r}}\times 
\mathbf{X_{\phi}}$ to the strip at every point.

We notice that this normal vector field is not continuous: if we 
compare the expressions at the center of the segment, $r=0$, after 
completing a turn from $\phi=0$ to $\phi=2\pi$,
\[ \mathbf{N}(0,0)=(0,0,1)\times (0,R,0)=(-R, 0,0),\]
\[\mathbf{N}(0,2\pi)=(0,0,-1)\times (0,R,0)=(R, 0,0),\]
the normal vector changes from pointing out of the center of the 
circumference to pointing towards the center, though the point on the 
strip is the same. Hence, the M\"obius strip is not orientable.

The boundary $\Gamma$ of the M\"obius strip $S$ is the curve described
by both endpoints $\{-a,a\}$ of the segment on rotating.  Or
equivalently, since the endpoint $a$ arrives at the original position
of $-a$ after a whole turn, we may describe $\Gamma$ by the motion of just
the endpoint $a$ after the segment travels twice along the
circumference to end up at the original position,
\[\gamma(\phi)=
\left
(\left(R+a\sin\frac{\phi}{2}\right)\cos\phi,
\left(R+a\sin\frac{\phi}{2}\right)\sin\phi,
a\cos\frac{\phi}{2}\right), 
\] for $\phi\in[0,4\pi]$.

\section{Flux \emph{across} a M\"obius' strip}

We are ready to perform some calculations on the strip and its
boundary.  For simplicity, we consider a simple constant vector field
along the $Z$ axis, $\mathbf{v}=(0,0,1)$. This field 
is solenoidal and a simple vector potential for it is 
$\mathbf{A}(x,y,z)=(-y/2,x/2,0)$. That is, 
$\mathbf{v}=\rot\,\mathbf{A}$.

The circulation of $\mathbf{A}$ along $\Gamma$, the boundary of the 
strip $S$ is well defined, since it is an oriented curve, and may be 
readily computed.

We need the velocity of the parametrisation of $\Gamma$,
 with velocity,
\begin{eqnarray*}\gamma'(\phi)&=&
\left
(-\left(R+a\sin\frac{\phi}{2}\right)\sin\phi,
\left(R+a\sin\frac{\phi}{2}\right)\cos\phi,
0\right)\\&+&\frac{1}{2}
\left
(a\cos\frac{\phi}{2}\cos\phi,
a\cos\frac{\phi}{2}\sin\phi,-
a\sin\frac{\phi}{2}\right), 
\end{eqnarray*}
and the  vector potential on the points of $\Gamma$ in this 
parametrisation,
\[\mathbf{A}(x(r,\phi),y(r,\phi),z(r,\phi))=\frac{1}{2}
\left(-\left(R+a\sin\frac{\phi}{2}\right)\sin\phi,
\left(R+a\sin\frac{\phi}{2}\right)\cos\phi,
0\right).
\]

Their inner product is just
\[\pe{\mathbf{A}(\gamma(\phi))}{\gamma'(\phi)}=\frac{1}{2}
\left(R+a\sin\frac{\phi}{2}\right)^{2},
\]
which makes the calculation of the circulation simple,
\begin{equation}\label{circA}\mathcal{C}_{\mathbf{A},\Gamma}=\int_0^{4\pi}\pe{\mathbf{A}(\gamma(\phi))}{\gamma'(\phi)}d\phi=
\frac{1}{2}\int_0^{4\pi}\left(R+a\sin\frac{\phi}{2}\right)^{2}d\phi=2\pi 
R^{2}+\pi a^{2}.\end{equation}

But if we naively calculate the flux of $\mathbf{v}$ across the strip,
\begin{eqnarray*}\Phi_{\mathbf{v},S}&=&\int_{-a}^{a}dr\int_{0}^{2\pi}d\phi
\pe{\mathbf{v}}{\mathbf{X_{r}}\times \mathbf{X_{\phi}}}=
\int_{-a}^{a}dr\int_{0}^{2\pi}d\phi
\left(R + r\sin\frac{\phi}{2}\right)\sin\frac{\phi}{2}
\\&=&8Ra,
\end{eqnarray*}
which of course does not provide the same result as the circulation 
of $\mathbf{A}$ along the boundary $\Gamma$, since the strip is not 
orientable and Stokes' theorem is not applicable.
\begin{figure}[h]
		\begin{center}
		    \includegraphics[height=0.2\textheight]{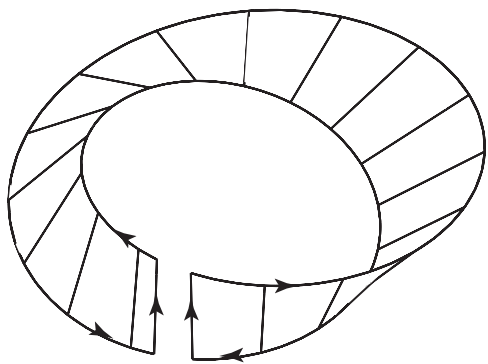}
\caption{Open M\"obius strip\label{broken}}
\end{center}
\end{figure}

However, there is a way to provide a meaning and an interpretation to the
previous integral.  If we cut the strip along the original segment at
$\phi=0$, we obtain an oriented open strip, but its boundary is not
$\Gamma$ as one could expect, but the union $\tilde \Gamma$ of four
pieces: the piece of $\Gamma$ corresponding to $\phi\in(0,2\pi)$, the
piece of $\Gamma$ corresponding to $\phi\in(2\pi,4\pi)$ with reversed
orientation and the original segment $I$ counted twice to link both
segments of $\Gamma$ (see Fig~\ref{broken}).  Since $I$ is orthogonal to $\mathbf{A}$, it
does not contribute to the circulation,
\begin{eqnarray*}\mathcal{C}_{\mathbf{A},\tilde\Gamma}&=&
\int_0^{2\pi}\pe{\mathbf{A}(\gamma(\phi))}{\gamma'(\phi)}d\phi
-\int_{2\pi}^{4\pi}\pe{\mathbf{A}(\gamma(\phi))}{\gamma'(\phi)}d\phi
\\&=&
\frac{1}{2}\int_0^{2\pi}\left(R+a\sin\frac{\phi}{2}\right)^{2}d\phi-
\frac{1}{2}\int^{4\pi}_{2\pi}\left(R+a\sin\frac{\phi}{2}\right)^{2}d\phi
=8Ra,\end{eqnarray*} 
which of course provides the same result as the
flux across the open strip, since Stokes'theorem is applicable to
this oriented surface. 

Though of course it is not the result we are 
after, since we wish to recover the circulation of $\mathbf{A}$ 
along $\Gamma$, not the flux of $\rot\mathbf{A}$ across a broken 
M\"obius strip.

\section{Circulation along the boundary of the strip}

We have checked explicitly that the flux of a solenoidal field 
across a M\"obius' strip and the circulation of its potential vector 
along the boundary of the strip are not the same, since Stokes' 
theorem cannot be applied to a non-orientable surface.

However, the circulation of the field along the boundary of the strip 
does have a physical meaning. As we have already mentioned, it could 
be the electromotiv force induced on a circuit located along $\Gamma$ 
by a varying magnetic field. Is it possible to calculate it using 
Faraday's law?

When written in this way we notice that the answer is simpler as 
expected when we formulated the question in terms of the flux across 
a M\"obius' strip, which sounded more appealing. Our goal is not the 
flux, which is an auxiliary quantity, but the circulation or 
electromotiv force, which is the one we can measure.

And again Stokes' theorem is of much help, since it can suggest the 
right answer to the right question. If we are interested in 
calculating the circulation $\mathcal{C}_{\mathbf{v},\Gamma}$, we 
notice that Stokes' theorem simply states that it can be done with 
the flux across \emph{any} oriented surface bounded by $\Gamma$. That 
is, M\"obius strip has $\Gamma$ as boundary and has been useful for 
defining it, but that is all: the strip is a bad choice, 
since it is not an oriented surface. But we can use any other 
oriented surface with the same boundary, as suggested in Exercise 
7.30 in \cite{berkeley}.

Cones are the simplest choice, since any closed curve without
self-intersections can be the boundary of a cone.  We take
any point $P$ in space as the vertex of the cone and draw the segments
that link $P$ with the points of $\Gamma$.  The resulting surface is a
cone bounded by $\Gamma$ and is an oriented surface.  The only issue
is that we have to choose $P$ so that the cone does not have
self-intersections.

\begin{figure}[h]
		\begin{center}
 \includegraphics[height=0.1\textheight]{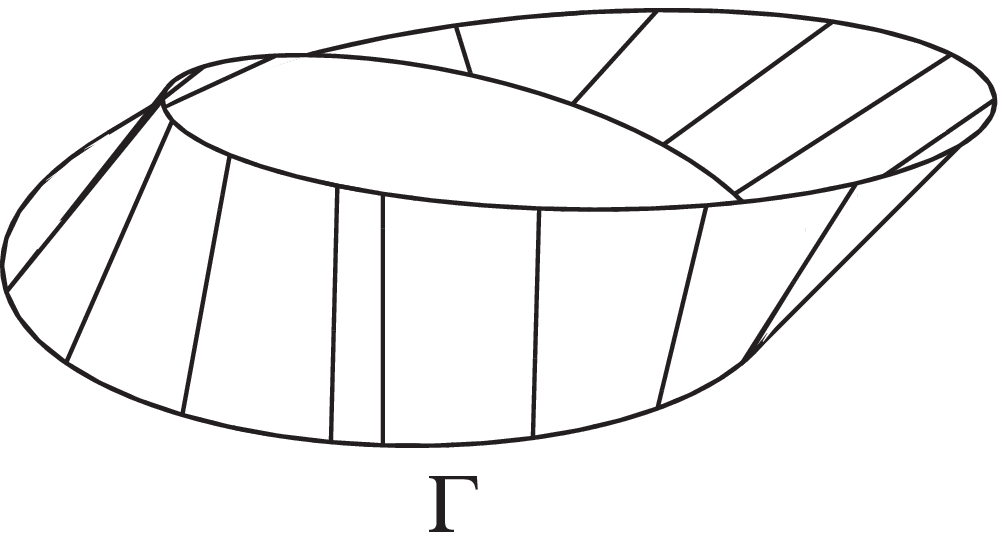}
			\includegraphics[height=0.1\textheight]{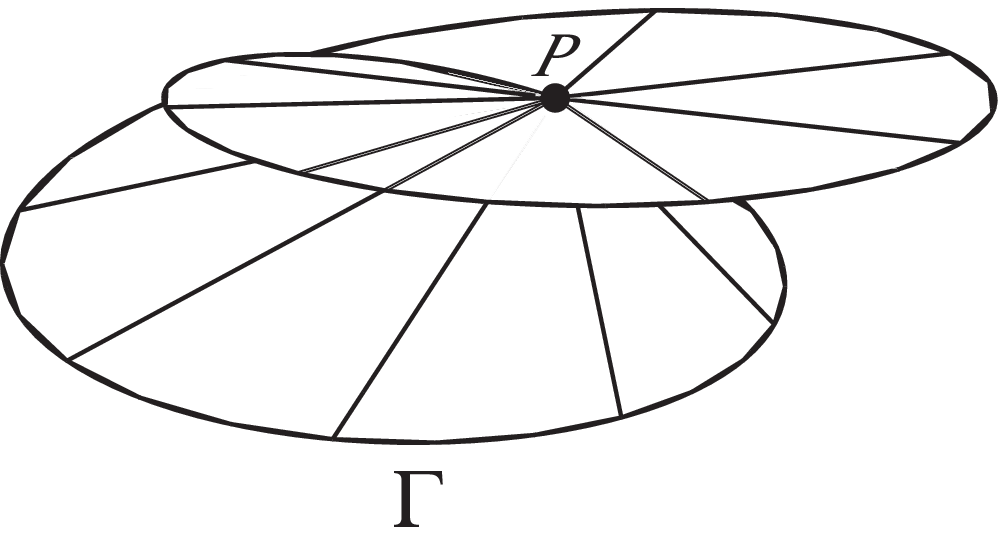}
\caption{M\"obius strip and orientable cone bounded by $\Gamma$\label{cone}: Both 
surfaces are bounded by the curve $\Gamma$. The cone is
constructed by linking the point $P$ with every point on $\Gamma$}
\end{center}
\end{figure}

A simple choice for the vertex is $(-R,0,0)$ (see Fig.~\ref{cone}), the
middle point of the horizontal segment on the strip at $\phi=\pi$.

A parametrisation for this cone $\tilde S$ is obtained by linear interpolation 
of a surface between the vertex, $\tilde g(0,\phi)$, and $\Gamma$, 
$\tilde g(1,\phi)$,
\[\tilde g(r,\phi)=(1-r)(-R,0,0)+ r\gamma(\phi), \qquad r\in(0,1), \ 
\phi\in(0,4\pi),\]
with the corresponding velocities for the coordinate lines,
\[\mathbf{X_{r}}(r,\phi)=\gamma(\phi)-(R,0,0),\qquad
\mathbf{X_{\phi}}(r,\phi)=r\gamma'(\phi),\]
allows calculation of the flux of $\mathbf{v}$ across the cone,
\begin{eqnarray*}\Phi_{\mathbf{v},\tilde 
	S}&=&\int_{0}^{1}dr\int_{0}^{4\pi}d\phi
\pe{\mathbf{v}}{\mathbf{X_{r}}\times \mathbf{X_{\phi}}}\\&=&
\int_{0}^{1}dr\int_{0}^{4\pi}d\phi
\left(2R^2\cos^{2}\frac{\phi}{2} 
 + Ra\left(1+3\cos^{2}\frac{\phi}{2}\right)\sin\frac{\phi}{2} + a^2\sin^{2}\frac{\phi}{2}
 \right)r
\\&=&2\pi R^{2}+\pi a^{2},
\end{eqnarray*}
and obtain the same results as with the circulation (\ref{circA}), 
according to Stokes' theorem, since the cone is orientable.

Calculations provide the same result for any other choice of the 
vertex $P$ of the cone.

\section{Conclusions}

In this paper we have provided a simple answer to the calculation of 
the flux of a vector field across a one-sided surface, where Stokes' 
theorem is not applicable.

We have shown that, though the question is ill posed, there is a way 
of restating the problem in order to provide a right answer, that is 
related to experiments we may perform in a laboratory.

It has been pointed out that the physically meaningful quantity is 
not the flux across the one-sided surface, but the circulation along 
the boundary of the surface. This quantity is not also meaningful, 
but can be measured, for instance, as the electromotiv force along a 
circuit induced by a varying magnetic field.

In fact, once we focus in computing the circulation along the
boundary, we notice that the one-sided surface is auxiliary and may be
replaced by \emph{any} other surface with the same boundary.  If the
chosen surface is orientable, this allows us to calculate the flux
and the circulation an obtain the same result, according to Stokes'
theorem. In fact, cones are always available for designing orientable 
surfaces with a given closed curve as boundary.

Summarising, the circulation of a vector field along the boundary of a
M\"obius strip, or any other one-sided surface, can be calculated
using Stokes' theorem, though not using the M\"obius strip, but any 
other surface with the same boundary.

\end{document}